\documentclass[%
aip,
twocolumn,
reprint,
10pt,
groupedaddress,
showpacs,
showkeys,
numerical,
footinbib
]{revtex4-1}

\newcommand*{\PZT}{Pb(Zr$_{0.2}$Ti$_{0.8}$)O$_3$}
\newcommand*{\SRO}{SrRuO$_3$}
\newcommand*{\STO}{SrTiO$_3$}
\newcommand*{\fref}[1]{Fig.~\ref{#1}}
\newcommand*{\frefs}[1]{Figs.~\ref{#1}}

\usepackage{graphicx}

\usepackage{color}

\usepackage[latin1]{inputenc}

\usepackage[bookmarks=false,pdfnewwindow,colorlinks,pdftitle={Minimum domain size and stability in carbon nanotube-ferroelectric devices},pdfauthor={Cedric Blaser and Patrycja Paruch}]{hyperref}
\hypersetup{citecolor=blue,linkcolor=blue,urlcolor=blue}

\begin{document}

\title{{\color{blue}Minimum domain size and stability in carbon nanotube-ferroelectric devices}}

\author{Cédric Blaser}
\email[Electronic mail: ]{Cedric.Blaser@unige.ch}
\affiliation{DPMC-MaNEP, University of Geneva, 24 Quai Ernest-Ansermet, 1211 Geneva 4, Switzerland}
\author{Patrycja Paruch}
\affiliation{DPMC-MaNEP, University of Geneva, 24 Quai Ernest-Ansermet, 1211 Geneva 4, Switzerland}
\date{Received 22 August 2012; accepted 24 September 2012; published online 5 October 2012}

\begin{abstract}
Ferroelectric domain switching in \textit{c}-axis-oriented epitaxial Pb(Zr$_{0.2}$Ti$_{0.8}$)O$_3$ thin films was studied using different field geometries and compared to numerical simulations and theoretical predictions. With carbon nanotubes as electrodes, continuous nanodomains as small as 9 nm in radius in a 270 nm thick film could be switched, remaining stable for over 20 months. Defect pinning of domain walls appears to play a key role in stabilizing such domains, below the predicted thermodynamic size limit.

\noindent{}\copyright{} \textit{2012 American Institute of Physics}. [\href{http://link.aip.org/link/doi/10.1063/1.4757880?ver=pdfcov}{http://dx.doi.org/10.1063/1.4757880}]
\end{abstract}
\maketitle

Control of nanoscale domain structures with different orientation of spontaneous electric polarization in ferroelectric thin films allows their integration into a wide range of applications\cite{scott_memories,kumar_apl_04_SAW}. From initial demonstrations of small domain size and high stability\cite{hidaka_afm_retention,maruyama_APL_98_arrays_FE}, through continued enhancement of domain density\cite{tybell_prl_02_creep,cho_apl_02_AFM_Tbit}, the local, high-intensity fields generated by metallic atomic force microscopy (AFM) tips were a key element. An important advance has been the functionalization of such tips with carbon nanotubes (CNTs), whose excellent metallic-state electrical conductivity, high strength, and small size have made them particularly interesting for polarization control. CNT bundles attached to AFM tips demonstrated domain switching in intermittent contact\cite{paruch_proc_02_AFM_CNT}, while subsequent use of CNT-AFM tips rigidified with SiO$_2$\cite{tayebi_apl_08_CNT_nanopencil} allowed full contact scanning, and both writing and imaging of domains as small as 2 nm in radius by piezoresponse force microscopy (PFM)\cite{tayebi_apl_10_CNT_PFM}.

From a fundamental perspective, understanding ferroelectric switching dynamics is intimately linked to the question of the smallest stable domain size. For a full description, many parameters, including electric field, film thickness, electrostatic/strain boundary conditions, and of course the presence of defects, need to be considered, both during and after switching. With efficient screening, nanoscale written domains can be thermodynamically stable. As domain size decreases, however, the effective force (line tension) of a high-curvature domain wall can promote collapse, especially in the presence of additional thermal activation\cite{paruch_prb_12_quench}. On the other hand, pinning of the domain wall on defects can provide a stabilizing mechanism\cite{tybell_prl_02_creep}. Understanding these effects is thus of clear applied, as well as fundamental interest. However, it is not trivial to explicitly include disorder in the mean-field models of domain walls in thermodynamic equilibrium. Experimental investigations of the switching and growth of domains in the limit of critical size, especially considering different electric field geometries and the long-term evolution of the domain structures, are therefore particularly useful.

Here, we report on such a study, using PFM to probe domain switching in three different electrode configurations, using the AFM tip, edges of macroscopic planar electrodes, and CNT on the ferroelectric surface as electric field sources, allowing us to separately consider the effects of the separation between two straight domain walls, and the radius of curvature on domain stability. In 270~nm thick films we demonstrate domain sizes down to 9~nm half-width with CNT and 14.5~nm radius with the AFM tip, two times smaller than the minimum domain size predicted from thermodynamic models\cite{morozovska_prb_06_AFM_domain,morozovska_prb_09_nanodomain_formation}. The observed domains remained stable over 20 months.

\begin{figure}[t]
\includegraphics{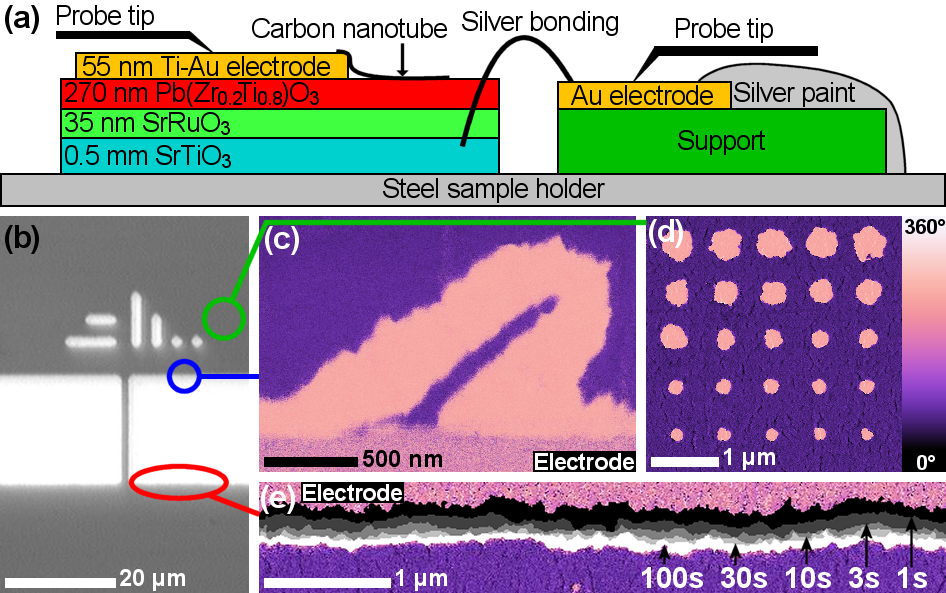}
\caption{\label{device}(a) Schematic side view of the device. (b) Central part of the device (1000x optical magnification), with the top electrodes and location markers. PFM phase images of ferroelectric domains generated by a CNT (c), an AFM tip (d) and the edge of an electrode (e).}
\end{figure}

We used a 270~nm thick epitaxial ferroelectric \PZT{} (PZT) thin film grown by off-axis radio frequency magnetron sputtering on 35~nm thick conducting \SRO{} on (001) single crystal \STO, as shown schematically in \fref{device}(a), with high crystalline and surface quality\cite{gariglio_apl_07_PZT_highTc}. The polarization axis is perpendicular to the film plane, with the film monodomain (up-polarized) as-grown. Ti(5~nm)/Au(50~nm) electrodes were deposited with standard photolithographic patterning and e-beam evaporation, and single-walled CNT subsequently dispersed on the ferroelectric surface from aqueous suspension \footnote{CNTs from \textit{SES Research} were suspended in deionized H$_2$O with sodium dodecylbenzenesulfonate surfactant \cite{islam_cnt_suspension}, then dispersed by spin-coating (one 20~$\mu$l drop, 3 min wait, 4000~rps for 30~s, 3 repeats)}, at a density of 0.05~CNT/$\mu$m$^2$. CNTs electrically contacting the electrodes were located by tapping mode topographic scans. As shown in \fref{device}(b), the devices can be used to study domain growth under such CNT, from the edge of the electrodes themselves, and under an AFM tip set down on the sample surface. To switch the polarization, voltage pulses were applied to the electrodes/CNT/AFM tip with the \SRO{} layer grounded, and the resulting domains imaged by PFM \footnote{Switching pulses to the electrodes/CNT were applied via ZN50R-10-BeCu needle probes (\textit{Lake Shore CPX} probe station, \textit{Agilent 81110A} pulse generator). Switching pulses to the \textit{Bruker MESP} tips were applied directly on the \textit{Veeco Dimension~V} AFM. For PFM imaging, typical parameters were 20~kHz drive frequency, 3000~mV drive amplitude and 3~$\mu$m/s tip velocity.}.

The single-walled CNTs used were both metallic and semiconducting in the standard 1:2 ratio. Since on ferroelectric perovskite surfaces non-metallic CNTs act as \textit{p}-type semiconductors \cite{paruch_apl_08_CNT_ferro}, all the CNT investigated could switch the intrinsic $P_{\rm UP}$ polarization to $P_{\rm DOWN}$ under a positive bias, and no difference in domain size was detected between the two CNT types. However, since a negative bias is necessary to reverse the $P_{\rm DOWN}$ polarization, only metallic CNT could be used for multiple polarization reversals such as that shown in \fref{device}(c). Moreover, as can be seen from the relative sizes of the $P_{\rm DOWN}$ vs. $P_{\rm UP}$ domains, both written with 1000~s pulses of +10~V and -10~V, respectively, domains written with positive bias appear to switch more easily. This asymmetric switching behavior, observed in our previous measurements \cite{paruch_prb_12_quench} and reported by other studies \cite{maksymovych_nanotech_11_PZT_conduction}, can be attributed to the asymmetric nature of the device itself, with different electrode work functions and geometries. For the quantitative studies, we therefore focused specifically on $P_{\rm DOWN}$ domains switched with positive applied bias. We note here that the PFM measurements were destructive for the CNT: the devices were therefore only imaged at the end of the writing process. However, for the domains written around the edges of the top electrodes, multiple PFM scans could be acquired, allowing progressive growth of domains with increasingly longer duration voltage pulses to be measured, as shown in the composite PFM image of \fref{device}(e). For the studies of radial domain growth, arrays of nanoscale circular domains were written by applying different duration voltage pulses to the AFM tip, as shown in \fref{device}(d).

\begin{figure}[t]
\includegraphics{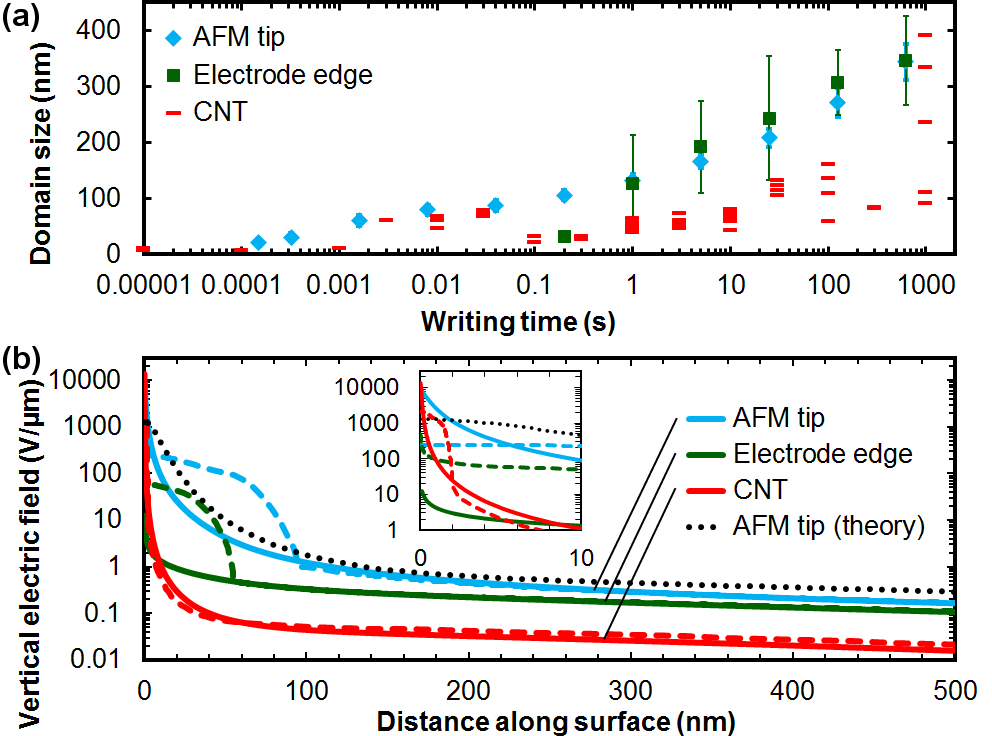}
\caption{\label{domain_dynamics}(a) Domain size as a function of writing time with +10~V bias. Error bars for AFM-tip and electrode-edge-written domains are the standard deviation of their dataset and may be smaller than the corresponding marker. Results for CNT-written domains are presented individually due to lower statistics. (b) Simulated vertical electric field as a function of distance along the surface from the field source for a 270~nm PZT thin film, with (dashed curves) and without (solid curves) a water meniscus, with inset on the short-distance region. The dotted curve is calculated following Ref. \protect\onlinecite{morozovska_prb_09_nanodomain_formation} for a generic AFM tip with an effective charge-surface separation of 10~nm.}
\end{figure}

To compare domain growth under the three different electrode geometries, we considered the radius and half-width, respectively, of the AFM-tip and CNT-written domains, growing symmetrically from the local electric field source, and the full width for the domains at electrode edges, since these grow uniformly around the entire electrode perimeter \footnote{For the AFM-written domains, the radius was extracted from the total domain area obtained from the binarized PFM phase image. For the CNT and electrode-edge-written domains, the average half-width/width was extracted from a linear portion of the domain.}. As shown in \fref{domain_dynamics}(a) for domains written with 10~V, 10~$\mu$s--1000~s pulses applied to the CNT, 10~V, 150~$\mu$s--625~s pulses applied to the AFM tip, and 10~V, 0.2~s--625~s pulses applied to the top electrodes, in all cases a logarithmic dependence of domain size on the writing time is observed, in agreement with our previous results \cite{tybell_prl_02_creep,paruch_apl_08_CNT_ferro}. An advantage of the present study is the use of undamaged epitaxial PZT, undeteriorated by direct growth of CNT\cite{paruch_apl_08_CNT_ferro}, allowing very short writing times to be explored. We find stable CNT-written domains for writing times down to 10~$\mu$s, but no evidence of switching for 3~$\mu$s, 10~V pulses. We note that these smallest domains are discontinuous, whereas domains written with pulses longer than 0.01~s are all fully continuous. With the AFM tip as field source, domains could be switched with pulses down to 150~$\mu$s, and using the electrode edge, down to 0.2~s. For longer writing times, we find that the widest domains grow outwards from the straight edges of the Ti/Au top electrodes, with slightly smaller domain radii obtained via AFM-writing, and still smaller domains under an overlying CNT. However, whereas AFM tip writing produces a relatively small spread of domain sizes for a given writing time, the size of domains written via electrode edges and CNT varies much more significantly.

\begin{figure}[t]
\includegraphics{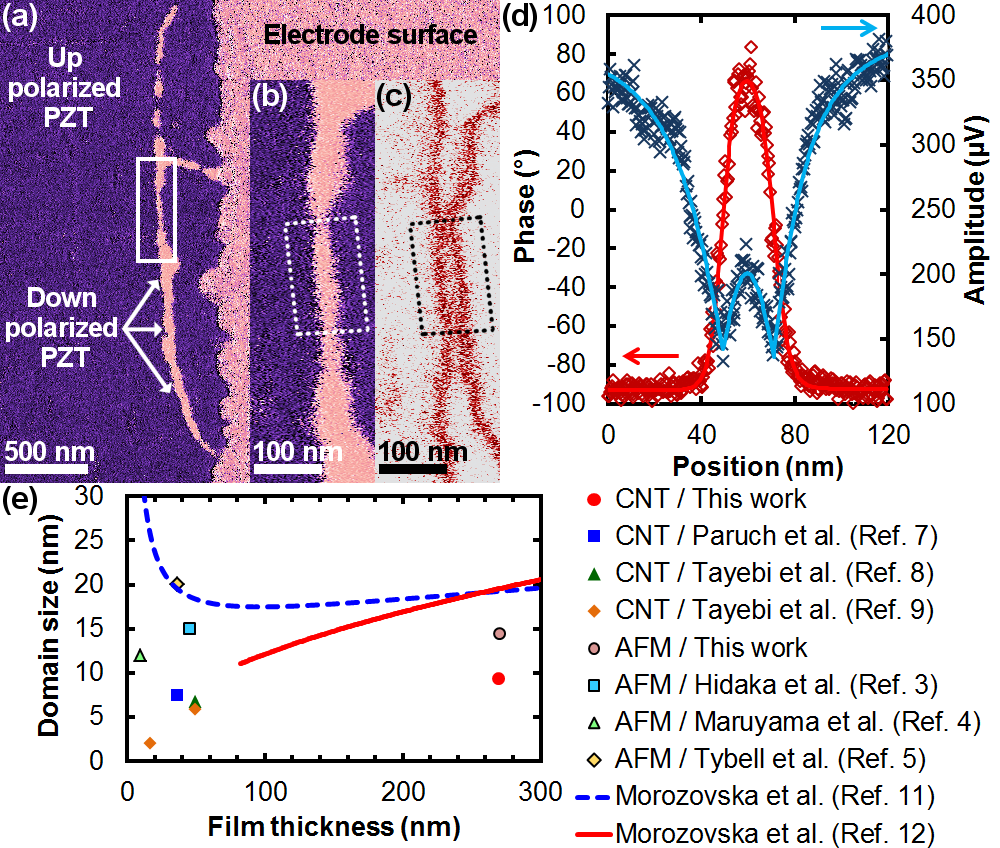}
\caption{\label{domain_size}(a) PFM phase image after a +10~V, 10~$\mu$s pulse, with the switched domain following the shape of the electrically connected CNT. Zoom on the thinnest continuous domain in (b) PFM phase and (c) PFM amplitude. (d) PFM phase and amplitude signals across the outlined area in (b)/(c). (e) Comparison of our data with previously reported domain sizes and models.}
\end{figure}

To better understand the observed results, we considered the vertical component of the electric field (along the polarization axis of the film) generated by the 10~V bias applied in each electrode configuration, numerically modeled using \textit{COMSOL Multiphysics} \footnote{In all three cases, we used a 2D model were the film is modeled by a 270~nm thick and 10~$\mu$m long rectangle with a relative dielectric permittivity of 80 (measured on the device). The AFM tip was modeled with a 10~$\mu$m high 21° half angle cone terminated by a 50~nm radius disc, the top electrode with a rectangle of 55~nm by 5~$\mu$m and the CNT with a 1~nm radius disc. The shape and the size of the water meniscus were adapted from Weeks \textit{et al.}\cite{weeks_langmuir_05_meniscus_shape_AFM_tip} and modeled by a Bézier polygon with the same contact angle on the field source and the film.}. Since all the polarization switching was carried out under ambient conditions in $\sim$45\% relative humidity, we considered this vertical electric field as a function of the horizontal distance along the film surface away from the field source, both with and without the effect of a water meniscus, shown by the dashed and solid lines in \fref{domain_dynamics}(b), respectively. In its immediate vicinity, the CNT generates the most intense vertical electric field, but this drops off very rapidly with horizontal distance, at 10~nm becoming less intense than either the field of the AFM tip or the electrode edge, as shown in the inset of \fref{domain_dynamics}(b). This high intensity and very rapid decrease agree well with the experimental observation that only using CNT could ferroelectric domains be switched with very short pulses down to 10~$\mu$s, but that attempts to write large domains with long pulses are inefficient compared to the other geometries. For the fields generated by the AFM tip and the electrode edges, our model suggests very similar fields at several hundreds of nanometers, and a higher AFM tip field close to the origin. This is in good qualitative agreement with the data, where we observe successful switching with much shorter pulses on the AFM tip than the electrode edge and similar domain sizes for pulse lengths between 1~s and 625~s, although those written with the electrode edge are slightly larger.

Since the key technological interest is minimizing the size of ferroelectric domains in a given device, we investigated in greater detail the behavior of CNT-written domains with the shortest writing times. Using 10~V, 10~$\mu$s voltage pulses, the smallest stable domains had half widths at half maximum as small as 9--13~nm, as shown in \fref{domain_size}(a), less than 4\% of the film thickness. Such small domains, given the convolution of the finite size of the tip (nominal radius of 50~nm, increasing on contact scanning) with the PFM signal, show a reduced amplitude signal compared to the bulk of the film. However, since we observe two PFM amplitude minima at the position of the domain walls, as shown in \frefs{domain_size}(b,c,d), we believe that these domains extend fully through the film. As described above, we observed no switching with shorter 10~V pulses. Increasing the voltage to 20~V did allow switching with 3~$\mu$s writing times, but the resulting domains were comparable to those obtained with the 10~V, 10~$\mu$s pulses, suggesting that a minimal domain radius of $\sim$10~nm can be written in the 270~nm film with overlying CNT.

\begin{figure}[t]
\includegraphics{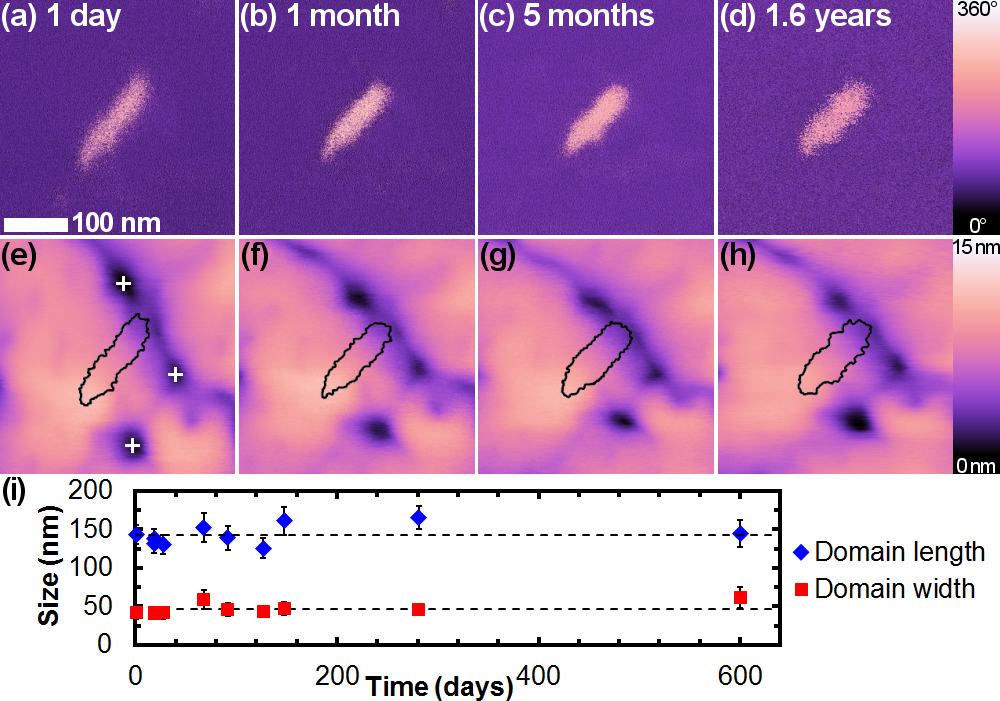}
\caption{\label{domain_stability}(a-d) PFM phase images of a CNT-written ferroelectric nanodomain generated by a +10~V, 100~$\mu$s pulse, after a time between 1 and 600 days after switching. (e-h) Simultaneously recorded topographic images of the same areas, with the outline of the ferroelectric nanodomain. (i) Length and width of the nanodomain as a function of time after topographic drift correction. The dashed lines are the respective averages.}
\end{figure}

Focusing on the smallest domains, we followed their evolution after the initial switching pulse. As shown in \fref{domain_stability} there is no significant size change even after more than 600 days. To take into account the effect of drift in AFM measurements, we rescaled the PFM phase images based on simultaneously recorded topographies, using the three local minima highlighted with a cross on \fref{domain_stability}(e). Of 37 domains considered, only 3 showed full backswitching, and 1 decreased in size over a few days, while the rest remained stable for the full observation period. Interestingly, for those domains showing backswitching, this apparently originated from the highly curved domain edges, decreasing the length of the domain, but not its width. These results suggest that in spite of their small size, the vast majority of the $\sim$10~nm radius domains we observe are indeed fully stabilized.

To compare our results with theoretical predictions of minimum domain size, we considered models describing domain switching under the highly localized electric field of an AFM tip, all of which focus on optimizing the energy balance between the nucleation and growth of a new volume of polarization parallel to the applied electric field, domain wall energy costs, depolarizing field at charged domain walls, as well as possible screening effects, either from surface adsorbates or free charges present in the film itself. From the earliest work \cite{kalinin_prb_02_pfm_FE}, adapting theories of switching in uniform fields \cite{landauer_jap_57_nucleation_BTO_electrostatics}, it was clear that a minimum switching voltage, depending on both tip (size, contact) and ferroelectric material parameters (film thickness, dielectric properties, screening) was necessary to nucleate a stable domain, remaining once the field was removed. In bulk materials and thicker films, such domains are prolate and do not necessarily penetrate through the sample, leading to depolarizing fields and broadened charged domain walls, while in thin films, rapid forward (vertical) growth of the domain across the sample is followed by slower lateral growth, giving a cylindrical domain shape (see Ref. \onlinecite{morozovska_prb_09_nanodomain_formation} and references therein for a detailed discussion). In the most complete thermodynamic models, using data from domain growth studies to extract the necessary material parameters like activation field, Debye screening radius and domain wall energy, minimal domain radii of 17.5~nm with critical voltage $U_{cr}$ = 7.1~V for a 97~nm thick PZT film, and 19.3~nm domain radius with $U_{cr}$ = 10.6~V for a 270~nm thick film like ours are expected (\fref{domain_size}(e)) \cite{morozovska_prb_06_AFM_domain,morozovska_prb_09_nanodomain_formation}. In contrast, the model used by Tayebi \textit{et al.}\cite{tayebi_apl_10_CNT_PFM}, based on domain stability considerations\cite{wang_jap_03_180DW_stability}, assumes a spherical domain shape, deemed ``unrealistic'' by Wang and Woo themselves, and a minimum size for the electrode necessary to stabilize a domain through the film thickness which would be orders of magnitude larger than our observations (where a 10~V bias applied to a CNT of 2~nm or less clearly allows stable domains to be generated through the 270~nm thick film).

The thermodynamic model of Morozovska \textit{et al.} is in excellent agreement with the experimental observations of critical switching voltage. However, even this model overestimates the minimum domain size by more than a factor of 2. The key difference is, we believe, the presence of defects, inherent in any real material, which can act as pinning and nucleation sites and stabilize domains below the limit expected from purely thermodynamic considerations in a perfect film. Furthermore, recent studies have shown that the intense, highly localized electric field of an AFM tip (or CNT) can lead to strong electrochemical reactions at the ferroelectric surface, in particular in ambient conditions, further modifying the defect landscape via surface charging, (re)ordering of oxygen vacancies, as well as more permanent damage at high voltages \cite{kalinin_nano_11_electrochemical_SPM}. In CNT overlying a ferroelectric surface, we had previously demonstrated strong, but ultimately reversible charging effects, whose relaxation could be followed via the CNT transconductance \cite{paruch_apl_08_CNT_ferro}. However, since the films are monodomain as-grown, we cannot quantify the relative importance of existing defects vs. the effects of those introduced during the domain writing process. In-situ switching studies in PZT have clearly demonstrated the pinning and slowing of domain walls by defects such as dislocations\cite{gao_natcom_11_PZT_TEM_switching}. In addition, during the initial stages of switching, the walls of the smallest domains incline, which has been linked with their increased conductivity\cite{maksymovych_nl_12_nanodomains_PZT}, compared to the relatively straight walls of larger domains\cite{guyonnet_am_11_DW_conduction}. Microscopically, such charged walls would interact even more strongly with defects such as oxygen vacancies.

In the present study, the discontinuous nucleation of small domains under the CNT and at the electrode edges demonstrates the range of nucleation bias across the film surface, and can be related to the presence of defects\cite{jesse_natmat_08_SSPFM}. While with an AFM tip, a single domain nucleates in the region of maximum electric field intensity, with both CNT and electrode edges, the high-field region extends linearly over larger portions of the film, allowing nucleation to occur at multiple sites and leading to greater variation in the final width of the resulting domains. Although the smallest domains show a tendency to collapse inwards from high-curvature domain walls, this can be clearly counteracted, and domains as small as 9~nm in radius readily stabilized in the relatively thick film for over 20 months. Finally, domains written with all three electrode configurations clearly show roughening, characteristic of the competition between domain wall elasticity and pinning\cite{paruch_prl_05_dw_roughness_FE}. These results suggest that a more detailed model taking into account domain wall pinning during switching, as well as experimental studies exploring the control of domain stability as a function of defect density and type would be extremely useful in the quest for ever smaller ferroelectric domains.

The authors thank S. Gariglio for the PZT samples used in this study, and M. Lopes and S. Muller for technical support. This work was funded by the Swiss National Science Foundation through the NCCR MaNEP and Division II grants 200021-121750 and 200020-138198.

%

\end{document}